\documentclass{article}

\usepackage{algorithm}
\usepackage{algorithmic}
\usepackage{geometry}
\usepackage{subfig}
\usepackage{amsmath}    
\usepackage{graphics}
\usepackage{graphicx}   
\usepackage{verbatim}   
\usepackage{color}      
\usepackage{graphics}
\usepackage{pdflscape}
\usepackage{setspace}
\usepackage{epsfig}
\usepackage[parfill]{parskip}    
\usepackage{url}
\usepackage[applemac]{inputenc}
\usepackage{nomencl}
\usepackage{amsfonts}
\usepackage{amsmath}
\usepackage{subfig}
\usepackage{multirow}
\usepackage{array}
\usepackage{colortbl}
\usepackage{hyperref}

\newcommand{\ket}[1]{|#1\rangle}

\let\oldsqrt\sqrt
\def\sqrt{\mathpalette\DHLhksqrt}
\def\DHLhksqrt#1#2{%
\setbox0=\hbox{$#1\oldsqrt{#2\,}$}\dimen0=\ht0
\advance\dimen0-0.2\ht0
\setbox2=\hbox{\vrule height\ht0 depth -\dimen0}%
{\box0\lower0.4pt\box2}}

\title{Quantum Iterative Deepening with an application to the Halting problem}

\author{
    Luís Tarrataca and Andreas Wichert\\
    Department of Informatics\\
    INESC-ID  / IST - Technical University of Lisboa\\
    Portugal\\
    \small{ \bf  \{luis.tarrataca,andreas.wichert\}@ist.utl.pt} 
}

\date{}

\begin{document}

\floatname{algorithm}{Procedure}

\maketitle

\begin{abstract}

The production system is a theoretical model of computation relevant to the artificial intelligence field allowing for problem solving procedures such as hierarchical tree search. In this work we explore some of the connections between artificial intelligence and quantum computation by presenting a model for a quantum production system. Our approach focuses on initially developing a model for a reversible production system which is a simple mapping of Bennett's reversible Turing machine. We then expand on this result in order to accommodate for the requirements of quantum computation. We present the details of how our proposition can be used alongside Grover's algorithm in order to yield a speedup comparatively to its classical counterpart. We discuss the requirements associated with such a speedup and how it compares against a similar quantum hierarchical search approach.


\end{abstract}

\floatstyle{ruled}
\newfloat{algorithmV2}{h}{lop}
\floatname{algorithmV2}{Algorithm }

\newcounter{DefinitionCounter}
\setcounter{DefinitionCounter}{1}

\newcounter{RequirementCounter}
\setcounter{RequirementCounter}{1}


\section{Introduction \label{sec:introduction}}

Classically, the status of any computation can be determined through a halt state. The concept of the halting state has some important subtleties in the context of quantum computation. The first one of these relates to quantum state evolution which needs to be expressed through unitary operators that represent reversible mappings. As a consequence, two successive states cannot be equal. Ekert draws attention to this fact stating that there are two possibilities to circumvent such an issue, namely \cite{ekert1996}: either run the computation for some predetermined number of steps or alternatively employ a halt flag. This flag is then employed by a computational model to signal an end of the calculation. Traditionally, such a flag is represented by a halt bit which is initialized to $0$ and set to $1$ once the computation terminates. Accordingly, determining if a computation has finished is simply a matter of checking if the halt bit is set to $1$, a task that can be accomplished through some form of periodic observation.

Furthermore, undecidable problems, such as the famous \textit{Entscheidungsproblem} challenge proposed by Hilbert in \cite{hilbert1900}, require that computational models be capable of proceeding indefinitely, a procedure that can only be verified through a recurrent observation of a halt bit. Classical models of computation are able to execute undecidable problems since their formulation allows for the use of such a flag without affecting the overall result of the calculation. Undecidable problems are important because they demonstrate the existence of a class of problems that does not admit an algorithmic solution no matter how much time or spatial resources are provided \cite{lewis1981}. This result was first demonstrated by Church \cite{church1936a} and shortly after by Turing \cite{turing1936}.

\subsection{Problem \label{sec:problem}}

Deutsch \cite{deutsch1985} was the first to suggest and employ such a strategy in order to describe a quantum equivalent of the Turing machine which employs a compound system $\ket{r}$ expressed as a tensor of two terms, \textit{i.e.} $\ket{r} = \ket{w}\ket{h}$, spanning a Hilbert space $\mathcal{H}_{r} = \mathcal{H}_{w} \otimes \mathcal{H}_{h}$. The component $\ket{w}$ represents a work register of unspecified length and $\ket{h}$ a halt qubit which is used in an analogous fashion to its classical counterpart. However, Deutsch's strategy turned out to be flawed, namely suppose a unitary computational procedure $C$ acting on input $\ket{x}$ is applied $d$ times and let $d_{C,x}$ represent the number of steps required for a procedure $C$ to terminate on input $x$. Then it may be possible that there exist $i$ and $j$ for which $d_{C,i} < d < d_{C,j}, \forall{i \neq j}$. Now, lets consider what happens when we are in the presence of such a behaviour and $\ket{w}$ is initialized as a superposition of the computational basis. Then those states which only require a number of computational steps less than or equal to $d$ in order to terminate will have the halt qubit set to $\ket{1}$, whilst the remaining states will have the same qubit set to $\ket{0}$.  This behaviour effectively results in the overall superposition state $\ket{w}\ket{h}$ becoming entangled as exemplified by Expression \ref{eq:deutschQuantumTuringMachineFlaw}, where we have assumed that $w$ employs $n$ bits.

\begin{equation}
\underbrace{\frac{1}{\sqrt{2^{n}}}\sum_{x = 0}^{2^{n} - 1} C^{d} \ket{ x }}_{\ket{\psi}}\ket{0} = 
	\begin{cases} 
	\ket{\underbrace{00 \cdots 0}_{\text{n bits}} } \ket{0} & \implies d_{C, 00 \cdots 0} > d\\
	\ket{00 \cdots 1 } \ket{1} & \implies d_{C, 00 \cdots 1} \leq d\\
	\quad \quad \vdots\\
	\ket{11 \cdots 0 } \ket{1} & \implies d_{C, 11 \cdots 0} \leq d\\
	\ket{11 \cdots 1 } \ket{0} & \implies d_{C, 11 \cdots 1} > d
	\end{cases}
\label{eq:deutschQuantumTuringMachineFlaw}
\end{equation}

More generally, suppose that the compound system after the unitary evolution $C^{d}$ is in the entangled state represented by the right-hand side of Expression~\ref{eq:compoundSystemBeforeMeasurement}. Also, assume that the probability of observing the halting qubit $\ket{h}$ with outcome $k$ is $P(k) = \sum_{x=0}^{2^{n}-1} |\alpha_{x,k}|^{2}$. The projection postulate implies that we obtain a post observation state of the whole system as the one illustrated in Expression \ref{eq:compoundSystemAfterMeasurement}, where the system is projected to the subspace of the halting register and renormalized to the unit length \cite{hirvensalo2004}.

\begin{equation}
\frac{1}{\sqrt{2^{n}}}\sum_{x = 0}^{2^{n} - 1} C^{d} \ket{ x } \ket{0} = \sum_{x=0}^{2^{n}-1} \sum_{j=0}^{1} \alpha_{x,j} \ket{x} \ket{j}
\label{eq:compoundSystemBeforeMeasurement}
\end{equation}

\begin{equation}
\frac{1}{\sqrt{P(k)}} \sum_{x=0}^{2^{n}-1} \alpha_{x,k} \ket{x} \ket{k}
\label{eq:compoundSystemAfterMeasurement}
\end{equation}

Consequently, observing the halt qubit after $d$ computational steps have been applied, will result in the working register containing either: (1) a superposition of the non-terminating states; or (2) a superposition of the halting states. Such behaviour has the 
 to dramatically disturb a computation since: (1) a halting state may not always be obtained upon measurement due to random collapse, if indeed there exists one; and (2) any computation performed subsequently using the contents of the working register $\ket{w}$ may employ an adulterated superposition with direct consequences on the interference pattern employed.  Roughly speaking, there is no way to know whether the computation is terminated or not without measuring the state of the machine, but, on the other hand, such a measurement may dramatically disturb the current computation.

\subsection{Current approaches to the quantum halting problem \label{sec:currentApproachesToTheQuantumHaltingProblem}}

Ideally, one could argue that any von Neumann measurement should only be performed after all parallel computations have terminated. Indeed, some problems may allow one to determine $\max d_{C,\ket{x}} ,\forall \ket{x} \in \ket{\psi}$, \textit{i.e.} an upper-bound $d_{C,x}$ on the number of steps required for every possible input $x$ present in the superposition. However, this procedure is not viable for those problems which, like the \textit{Entscheidungsproblem}, are undecidable. Bernstein and Vazirani subsequently proposed a model for a universal quantum Turing machine in \cite{bernstein1993} which did not incorporate into its definition the concept on non-termination. Although their model is still an important theoretical contribution it is nonetheless only capable of dealing with computational processes whose different branches halt simultaneously or fail to halt at all. These same arguments were later employed by Myers in \cite{myers1997} who argues that it is not possible to precisely determine for all functions that are Turing-computable, respectively $\mu$-recursive functions, the number of computational steps required for completion. Additionally, the author also states that the models presented in \cite{deutsch1985} and \cite{bernstein1993} cannot be qualified as being truly universal since they do not allow for non-terminating computation. The work described in \cite{bernstein1993} is also restricted to the class of quantum Turing machines whose computational paths are synchronized, \textit{i.e.} every computational path is synchronized in the sense that they must each reach an halt state at the same time step. This enabled the authors to sidestep the halting problem.

Following Myers observation of the conflict between quantum computation and system observation  a number of authors provided meaningful contributions to the question of halting in quantum Turing machines. Ozawa \cite{ozawa1998a} \cite{ozawa1998b} proposed a possible solution based on quantum nondemolition measurements, a concept previously employed for gravitational wave detection. Linden \cite{linden1998} argued  that the standard halting scheme for Turing machines employed by Ozawa is unitary only for non-halting computations. Additionally, the author described how to build a quantum computer, through the introduction of an auxiliary ancilla bit that enabled system monitoring without spoiling the computation. However, such a scheme introduced difficulties regarding different halting times for different branches of computation. These restrictions essentially rendered the system classical since no useful interference occurred. In \cite{ozawa2002} expands the halting scheme described in \cite{ozawa1998a} in order to introduce the notion of a well-behaved halting flag which is not modified upon completion. The author showed that the output probability distribution of monitored and non-monitored flags is the same. Miyadera proved that no algorithm exists capable of determining if an arbitrarily constructed quantum Turing machine halts at different computational branches \cite{miyadera2003}. Iriyama discusses halting through a generalized quantum Turing machine that is able to evolve through states in a non-unitary fashion \cite{iriyama2004}. 

Measurement-based quantum Turing machines as a model for computation were defined in \cite{perdrix2004a} and \cite{perdrix2004b}. Perdrix explores the halting issue by introducing classically-controlled quantum Turing machines \cite{perdrix2006}, in which unitary transformations and quantum measurements are allowed, but restricts his model to quantum Turing machines that halt. Muller shows the existence of a universal quantum Turing machine that can simulate every other quantum Turing machine until the simulated model halts which then results in the universal machine halting with probability one \cite{muller2007,muller2008}. The author describes operators that do not disturb the computation as long as the original input employed halts the calculation process. This requires presenting a precise definition of the concept of halting state. This notion results in a restriction where large parts of the domain are discarded since the definition requirements are not met. 

In \cite{ross2010} a method is presented for verifying the correctness of measurement-based quantum computation in the context of the one-way quantum computer described in \cite{raussendorf2001}. This type of quantum computation differs from the traditional circuit based approach since one-qubit measurements are performed on an entangled resource labeled as a cluster state in order to mold a quantum logic circuit on the state. With each measurement the entanglement resource is further depleted. These results are further extended in \cite{raussendorf2003} in order to prove the universality of the computational model. Subsequently, in \cite{browne2011} these concepts were used in order to prove that one-way quantum computations have the same computational power as quantum circuits with unbounded fan-out. Perdrix \cite{perdrix2011} discusses partial observation of quantum Turing machines which preserve the computational state through the introduction of a weaker form of the original requirements of linear and unitary $\delta$ functions  suggested by Deutsch in \cite{deutsch1985}. Recently, \cite{mhalla2012} proved that measurements performed on the $(X, Z)$-plane of the Bloch sphere over graph states is a universal measurement-based model of quantum computation.

\subsection{Objectives \label{sec:objectives}}

In its seminal paper \cite{deutsch1985}, Deutsch emphasizes that a quantum computer needs the ability to operate on an input that is a superposition of computational basis in order to be ``fully quantum'', When confronted with the halting issue Myers naturally raised the question if a universal quantum computer could ever be fully quantum? And how would such a computational model eventually function? We aim to provide an answer to these questions by developing an alternative proposal to quantum Turing machines based on production system theory. We introduce such a computational model in order to gain additional insight into the matter of halting and universal computation from a different perspective than that of the standard quantum Turing machine. 

As Miyadera stated, the notion of probabilistic halting in the context of quantum Turing machines cannot be avoided, suggesting that the standard halting scheme of traditional quantum computational models needs to be reexamined \cite{miyadera2003}.  Our proposal is essentially different from the ones previously discussed since it imposes a strict notion of how the computation is performed and progresses in the form of the sequence of instructions that should be applied. Our method evaluates $d$-length sequences of instructions representing different branches of computation, enabling one to determine which branches, if they exist, terminate the computation. Underlying the proposed model will be Grover's algorithm in order to amplify the amplitude of potential halting states, if such states exist, and thus avoiding obtaining a random projection upon measurement. As a result, we will focus on characterizing the computational complexity associated with such a model and showing that it does not differ from that of Grover's algorithm. 

With this work we are particularly interested in: (1) preserving the original principles proposed by Deutsch of linearity and unitary operators, in contrast with other proposals such as \cite{perdrix2011} and \cite{iriyama2004} which perform modifications to the underlying framework; (2) developing a model which considers all possible computational paths and (3) works independently of whether the computation terminates or not taking into account each possible computational path. Additionally, we will also consider some of the implications of being able to circumvent the halting problem. Computation universality is a characteristic attribute of several classical models of computation. For instance, the Turing machine model was shown to be equivalent in power to lambda calculus and production system theory. Accordingly, it would be interesting to determine what aspects of such a relationship are maintained in the context of quantum computation. Namely,  we are interested in determining if it is possible to simulate a classical Turing machine given a quantum production system.

\subsection{Organisation \label{sec:organisation}}

The ensuing sections are organised as follows: Section \ref{sec:productionSystemReview} presents the details of production system theory, a computational model that will be employed to model tree search applied to the halting problem; Section \ref{sec:QuantumIterativeDeepening} extends these ideas to a quantum context and discusses the details associated with our proposal for detection of quantum halting states. Section \ref{sec:turingMachineSimulation} demonstrates how our proposal can be employed in order to coherently simulate a classical Turing machine. We present the conclusions of this work in Section \ref{sec:conclusions}.

\section{Production System Review \label{sec:productionSystemReview}}

Our  approach  to  the  detection  of  quantum  halting  states  requires  fixing  a  computational  model.  This  step  is required since our proposal depends on the set of state transitions occurring during a computational process. We choose not to focus on Turing machines, instead our proposal will be formulated in terms of production system theory. This decision is based on the fact that the quantum Turing machine model was already well explored by Deutsch \cite{deutsch1985} as well as Bernstein and Vazirani \cite{bernstein1993}. Furthermore, the combination of quantum concepts such as interference, entanglement and the superposition principle alongside the halting issue also contribute to make these models inherently complex. As a result, it is difficult to express elementary computational procedures. This behaviour contrasts with the simplicity of production system theory which allows for an elegant and compact representation of computations.

Production system theory is also well suited to support tree search, a form  of  graph search from which we drew our initial inspiration. In addition, the classical counterparts of both models were shown to be equivalent in computational power \cite{abramsky1999}. The production system is a formalism for describing the theory of computation proposed by Post in \cite{post1943}, consisting of a set of production rules $R$, a control system $C$ and a working memory $W$. This sections reviews some of the most significant definitions that were proposed in \cite{tarrataca2011b}, namely:

\begin{description}

	\item[Definition \theDefinitionCounter \stepcounter{DefinitionCounter}] Let $\Gamma$ be a finite nonempty set whose elements are referred to as symbols. Additionally, let $\Gamma^{*}$ be the set of strings over $\Gamma$.

	\item[Definition \theDefinitionCounter \stepcounter{DefinitionCounter}] The working memory $W$ is capable of holding a string belonging to $\Gamma^{*}$. The working memory is initialized with a given string, who is also commonly referred to as the initial state $\gamma_{i}$.
	
	\item[Definition \theDefinitionCounter \stepcounter{DefinitionCounter}] The set of production rules $R$ has the form presented in Expression \ref{ch:finalConsiderations:eq:productionRule}.
	
	\begin{equation}
	\{(precondition, action) | precondition, action \in \Gamma^{*} \}
	\label{ch:finalConsiderations:eq:productionRule}
	\end{equation}
	
	Each rules precondition is matched against the contents of the working memory. If the precondition is met then the action part of the rule can be applied, changing the contents of the working memory. 
	
	\item[Definition \theDefinitionCounter \stepcounter{DefinitionCounter}] The tuple $(\Gamma, S_{i}, S_{g}, R, C )$ represents the formal definition of a production system where $\Gamma, R$ are finite nonempty sets and $S_{i},S_{g} \subset \Gamma^{*}$ are, respectively, the finite sets of initial and goal states. The control function $C$ satisfies Expression \ref{eq:productionSystemTuple}.
	
	\begin{equation}
	C : \Gamma^{*} \rightarrow R \times \Gamma^{*} \times \{h,c\}
	\label{eq:productionSystemTuple}
	\end{equation}
	
	The control system $C$ chooses which of the rules to apply and terminates the computation when a goal configuration, $\gamma_{g}$, of the memory is reached. If $C( \gamma) = (r, \gamma^{\prime}, \{h,c\})$ the interpretation is that, if the working memory contains  string $\gamma$ then it is substituted by the action $\gamma^{\prime}$ of rule $r$ and the computation either continues, $c$, or halts, $h$. Traditionally, the computation halts when a goal state $\gamma_{g} \in S_{g}$ is achieved through a production, and continues otherwise.
	
	\item[Definition \theDefinitionCounter \stepcounter{DefinitionCounter}] Let $\zeta_{d}$ represent a sequence of productions leading up to a state $s$ of length $d$. If $s \in S_{g}$ then such a sequence is also referred to as a solution.
		 	
\end{description}

\begin{figure}[ht]
\centering
\includegraphics[width=0.58\columnwidth ]{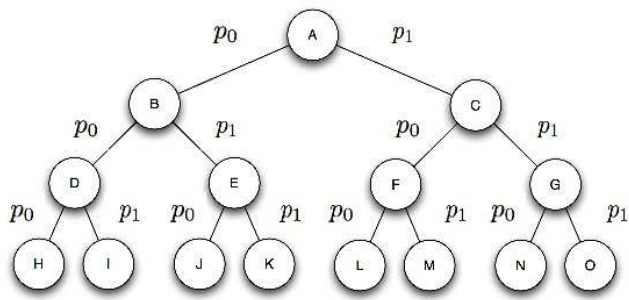}
\caption{Tree structure representing the multiple computational paths of a probabilistic production system. \label{fig:probabilisticProductionSystem}}
\end{figure}

Figure \ref{fig:probabilisticProductionSystem} illustrates a production system with two production rules namely $\{p_{0}, p_{1}\}$ that can always be applied. Thus the representation as a graph with a tree form, representing a search of depth level $3$ with initial state is $A$ and leaf $\{H, I, J, K, L, M, N, O\}$. Each depth layer $d$ adds $b^{d}$ nodes to the tree, where $b$ is the branching factor resulting from $|R|$, with each requiring a unique path leading to them. Therefore a total of $b^{d}$ possible paths exist, \textit{e.g.} state $J$ is achieved by applying sequence $\{p_{0}, p_{1}, p_{0}\}$. 

With these definitions in mind it becomes possible to develop a suitable model for a quantum production system. Namely, the complex valued control strategy would need to behave as illustrated in Expression \ref{eq:complexValuedControlSystem1} where $C( \gamma, r, \gamma^{\prime}, d)$ provides the amplitude if the working memory contains string $\gamma$ then rule $r$ will be chosen, substituting string $\gamma$ with  $\gamma^{\prime}$ and a decision $s$ made on whether to continue or halt the computation.

\begin{equation}
	C : \Gamma^{*} \times R \times \Gamma^{*} \times \{h,c\} \rightarrow \mathbb{C}
	\label{eq:complexValuedControlSystem1}
\end{equation}

The amplitude value provided would also have to be in accordance with Expression \ref{eq:complexValuedControlSystem1Normalization1}, $\forall \gamma \in \Gamma^{*}$ 

\begin{equation}
	\sum_{ \forall(r, \gamma^{\prime}, s)  \in R \times \Gamma^{*} \times \{h,c\} } |C( \gamma, r, \gamma^{\prime}, s )|^{2} = 1 
	\label{eq:complexValuedControlSystem1Normalization1}
\end{equation}

We will employ the notation described in \cite{hirvensalo2004} to describe the evolution of our quantum production system. Suppose we have a unitary operator $C$ with the form presented in Expression \ref{eq:complexValuedControlSystem1}. Operator $C$ is responsible for a discrete state evolution taking the system from state $\gamma$ to $\gamma^{\prime}$ through production $r$, expressed as $\gamma \vdash_{r} \gamma^{\prime}$. We refer to the transition $\gamma \vdash_{r} \gamma^{\prime}$ as a \textit{computational step}.  The computation of a production system starting in an initial state $i \in S_{i}$ can be defined as a sequence of steps $c_{1}, c_{2}, \cdots, c_{d}$ such that $c_{k} \vdash c_{k+1} \forall_{k}$ and where $d \in \mathbb{N}$ represents the depth at which a solution state $g \in S_{g}$ can be found. In general, the unitary operator $C$ can be perceived as applying a single computational step of the control strategy for a general production system. This notation is convenient since we are able to express the computation of a production system  $C$ up to depth-level $d$ as $C^{d}$, \textit{i.e.} a depth-limited search mechanism that mimics the behaviour illustrated in Figure~\ref{fig:probabilisticProductionSystem}.

\section{Quantum Iterative Deepening \label{sec:QuantumIterativeDeepening}}

Universal models of computation are capable of calculating $\mu$-recursive functions, a class of functions which allow for the possibility of non-termination. These functions employ a form of unbounded minimalization, respectively the $\mu$-operator, which is defined in the following terms \cite{lewis1981}: let $k \geq 0$, $c \in \mathbb{N}$,$m \in \mathbb{N}$  and $g \colon \mathbb{N}^{k + 1} \to \mathbb{N}$, then the unbounded minimization of $g$ is function $f \colon \mathbb{N}^{k+2} \to \mathbb{N}$ as illustrated in Expression \ref{eq:muOperator1}, for any $\bar{n} = n_{1}, \cdots, n_{k} \in \mathbb{N}^{k}$. 

\begin{equation}
f( g, \bar{n}, c )  = \left\{
\begin{array}{ll}
\text{the least $m$ such that $g(\bar{n}, m) = c$}  	& \text{, if such an $m$ exists}\\
0 										& \text{, otherwise}
\end{array} \right.
\label{eq:muOperator1}
\end{equation}

The unbounded minimization operator can be perceived as a computational procedure responsible for repeatedly evaluating a function with different inputs $m$ until a target condition $g(\bar{n}, m) = c$ is obtained \cite{stuart2004b}. However, as illustrated by Expression \ref{eq:muOperator1}, there is no guarantee that the target condition will ever be met.  Accordingly, it is possible to express the inner-workings of $f$ as an iterative search that may never terminate, as illustrated in Algorithm~\ref{algo:muOperator}. Notice that although $\mu$-recursive functions employ a collections of variables belonging to the set of natural numbers, for practical purposes these values are restricted by architecture-specific limits on the number of bits available for representing the range of possible values.

\begin{algorithmV2}
\begin{algorithmic}[1]

\STATE \textbf{function }$f( g, \bar{n}, c$)

\STATE $ \qquad m \leftarrow 0$

\STATE $ \qquad \text{\textbf{while }} g( \bar{n}, m) \neq c \textbf{ do}$

\STATE $ \qquad \qquad m \leftarrow m+1$

\STATE $ \qquad \text{ \textbf{return }} m$

\end{algorithmic}
\caption{The classical $\mu$-operator (adapted from \cite{stuart2004b})}
\label{algo:muOperator}
\end{algorithmV2}

From a quantum computation perspective, it is possible to perform a generic search for solution states through amplitude amplification schemes such as the one described by Grover in \cite{grover1996} and \cite{grover2004a}. In this section we will discuss how to combine  production system theory alongside the quantum search algorithm in order to develop a new computational model better suited to deal with the halting issue.

The next sections are organized in the following manner: Section \ref{sec:groverAlgorithm} presents the main details associated with Grover's algorithm; Section \ref{sec:quantumProductionSystemOracle} proposes an oracle formulation of a the quantum production; Section \ref{sec:generalProcedure} focuses on how to integrate these components into a single unified approach for a computational model based on production system theory capable of proceeding indefinitely without affecting the overall result of the computation; Section \ref{sec:turingMachineSimulation} presents a simple mapping mechanism of how our approach can be used to simulate a classical Turing machine.

\subsection{Grover's algorithm \label{sec:groverAlgorithm}}

The quantum search algorithm employs an oracle $O$ whose behaviour can be formulated as presented in Expression \ref{eq:oracle}, where $\ket{w}$ is a $n$-qubit query register, $\ket{h}$ is a single qubit answer register. Additionally, $f(w)$ is responsible for checking if $w$ is a solution to a problem, outputting value $1$ if so and $0$ otherwise. In the context of this research we only consider deterministic functions.

 \begin{equation}
	 O : \ket{ w } \ket{ h } \mapsto \ket{ w } \ket{h \oplus f( w ) }
 \label{eq:oracle}
 \end{equation}
 
 It is important to mention that we employed some care when defining the oracle in terms of registers $\ket{w}$ and $\ket{h}$, in a similar manner to the quantum Turing machine model proposed by Deutsch. We deliberately chose to do so in order to establish some of the connections between the halting problem and the quantum search procedure. We may view the halting problem as one where we wish to obtain the computational basis present in $\ket{w}$ which lead to goal states $g \in S_{g}$ where $S_{g}$ is defined as the set of halting states.
 
Grover's algorithm starts by setting up a superposition of $2^{n}$ elements in register $\ket{w}$ and subsequently employs a unitary operator $G$ known as Grover's iterate \cite{kaye2007a} in order to amplify the amplitudes of the goal states and diminish those of non-goal states. The algorithm is capable of searching the superposition of $2^{n}$ elements by invoking the oracle $O(\sqrt{2^{n}})$ times. The computational complexity of $f$ should also be taken into consideration. Namely, assume that $f$ takes time $t_{f}$. Since Grover's algorithm performs $\sqrt{2^{n}}$ oracle invocations then the  total complexity will be $O(\sqrt{2^{n}} t_{f})$. This complexity still represents a speedup over an equivalent classical procedure since $2^{n}$ states would have to be evaluated independently. However, for a polynomial $t_{f}$ the overall complexity will be dominated by the dimension of the search space, \textit{i.e.} $O(\sqrt{2^{n}})$. For this reason, it is often assumed that $f$ is computable in polynomial time. This assumption also makes such oracle models suitable to the complexity class NP which represents the class of languages that can be verified by a polynomial-time algorithm.

In addition it is also possible that the space includes several solutions. Accordingly, let $k$ represent the number of solutions that exist in the search space, then the complexity of the quantum search algorithm can be restated as $O\left( \sqrt{\frac{2^{n}}{k}} \right)$. Typically, $k$ can be determined through the quantum counting algorithm described in \cite{brassard2000} which also requires a similar time complexity. This means that before applying Grover's algorithm one must first determine the number of solutions. Overall, the time complexity of applying both methods sequentially remains the same. Once the algorithm terminates and a measurement is performed then a random collapse occurs, with high probability, amongst the amplified solutions. In the remainder of this work we gain generality by thinking in terms of the worst-case scenario where a single solution exists. However, the method described above could still be applied to the proposition that is described in the following sections. Grover's algorithm was experimentally demonstrated in~\cite{chuang1998}. 

\subsection{Quantum Production System Oracle \label{sec:quantumProductionSystemOracle}}

Is it possible to present an adequate mapping of our quantum production system that is suitable to be applied alongside Grover's algorithm? A comparison of Expression \ref{eq:complexValuedControlSystem1} and Expression \ref{eq:oracle} allows us to reach the conclusion that oracle $O$ performs a verification whilst $C$ focuses on executing an adequate state evolution. Therefore, we need to develop an alternate mechanism that behaves as if performing a verification. We can do so by focusing on one of the main objectives of production system theory, namely that of determining the sequence of production rules leading up to a goal state. Formally, we are interested in establishing if an initial state $i \in S_{i}$ alongside a sequence of $d$ productions rules $\{r_{1}, r_{2}, \cdots, r_{d}\} \in R$ leads to a goal state $g \in S_{g}$. If the sequence of rules leads to a goal state, then the computation is marked as being in a halt state $h$, otherwise it is flagged to continue $c$. We can therefore proceed with a redefinition of the control function presented in Expression \ref{eq:complexValuedControlSystem1}, as illustrated in Expression \ref{eq:complexValuedControlSystem2}, which closely follows the oracle definition presented in Expression~\ref{eq:oracle}. 

\begin{equation}
	C : \Gamma^{*} \times R^{d} \times \{h,c\} - \mathbb{C}
	\label{eq:complexValuedControlSystem2}
\end{equation}

Recall that the oracle operator is applied to register $\ket{r} = \ket{w}\ket{h}$. We choose to represent register $\ket{w}$ as a tensor of two products, namely $\ket{w} = \ket{s} \ket{p}$, where $\ket{s}$ is responsible for holding the binary representation of the initial state and $\ket{p}$ contains the sequence of productions. Register $\ket{h}$ is utilized in order to store the status $s$ of the computation. Additionally, the revised version of the quantum production system $C$ with oracle properties should also maintain a unit-norm, as depicted by Expression \ref{eq:complexValuedControlSystem1Normalization2}, $\forall \gamma \in \Gamma^{*}$. For specific  details surrounding the construction of such a unitary operator please refer to \cite{tarrataca2011c}.

\begin{equation}
	\sum_{ \forall(r_{1}, r_{2}, \cdots, r_{d}, s)  \in R^{d} \times \{h,c\} } |C( \gamma, r_{1}, r_{2}, \cdots, r_{d}, s )|^{2} = 1 
	\label{eq:complexValuedControlSystem1Normalization2}
\end{equation}

Any computational procedure can be described in production system theory by specifying an appropriate set of production rules that are responsible for performing an adequate state evolution. This set of production rules can be applied in conjunctions with a unitary operator $C$ incorporating the behaviour mentioned in Expression \ref{eq:complexValuedControlSystem2} and Expression \ref{eq:complexValuedControlSystem1Normalization2}. In doing so we are able to obtain a derivation of a production system that can be combined with Grover's algorithm. From a practical perspective, we are able to initialize $\ket{p}$ as a superposition over a set $P_{R,d}$ representing the sequence of all possible production rules $\in R$ up to a depth-level $d$, as illustrated by Expression \ref{eq:sequencePossibleProductions} and Expression \ref{eq:superpositionOfProductionRules}. Implicit to these definitions is the assumption that set $P$ has a total of $b^{d}$ possible paths.

\begin{equation}
P_{R,d} := \{\text{sequence of all possible production rules $\in R$ up to a depth-level } d\}
\label{eq:sequencePossibleProductions}
\end{equation}

\begin{equation}
\ket{p} = \frac{1}{\sqrt{b^{d}}} \sum_{\forall x \in P_{R,d}} \ket{x}
\label{eq:superpositionOfProductionRules}
\end{equation}

Traditionally, throughout a computation set $S_{i}$ remains static in the sense that it does not grow in size. However, variable $d$ is constantly increased in order to generate search spaces covering a larger number of states. As a result, given a sufficiently large depth value the number of bits required for $P_{R,d}$ will eventually surpass the amount of bits required to encode set $S_{i}$. Accordingly, in the reasonable scenario where the number of bits required to encode the sequence of productions over $P_{R,d}$ is much larger than the number of bits required to encode the set of initial states $S_{i}$, \textit{i.e.} $\log_{2}{|P_{R,d}|} \gg \log_{2}{|S_{i}|}$, then the most important factor to the dimension of the search space will be the number of productions. For this reason, Grover's algorithm needs to evaluate a search space spanning roughly a total of $b^{d}$ paths. As a consequence, the algorithm's running time is $O(\sqrt{b^{d}})$ which effectively cuts the search depth in half \cite{tarrataca2010}.

\subsection{General procedure \label{sec:generalProcedure}}

Any approach to a universal model of quantum computation needs to focus on two main issues, namely: (1) how to circumvent the halting problem and (2) how to handle computations that do not terminate without disturbing the result of the procedure. In the next sections we describe our general procedure. We choose to focus first on the second requirement in Section \ref{sec:parallelsBetweenMuTheoryAndProductionSystemTheory} given that it provides a basis for model development by establishing the parallels between $\mu$-theory and production system theory. We then describe in Section \ref{sec:iterativeSearch} how these arguments can be utilized in order to develop a computational model capable of calculating $\mu$-recursive functions. We conclude with Section \ref{sec:complexityAnalysis} where we describe how our proposal is essentially non-different, complexity-wise, from the original Grover algorithm employed thus allowing for an efficient method satisfying both requirements.

\subsubsection{Parallels between $\mu$-theory and production system theory \label{sec:parallelsBetweenMuTheoryAndProductionSystemTheory}}

Universal computation must allow for the possibility of non-termination, a characteristic that is is achievable through the ability to calculate $\mu$-recursive functions. Therefore, the question naturally arises if it is possible to develop a quantum analogue of the iterative $\mu$-operator? By itself $\mu$-recursive functions are not seen as a model of computation, but represent a class of functions that can be calculated by computational models. Accordingly, we are interested in determining if we are able to develop a quantum computational model, namely by employing the principles of production system theory, capable of calculating $\mu$-recursive functions without affecting the end result. 

In order to answer this question we will first start by establishing some parallels between these concepts. Namely, consider the $\mu$-operator presented in Algorithm \ref{algo:muOperator} that receives as an argument a tuple $(g, \bar{n}, c)$ and a production system defined by the tuple $(\Gamma, S_{i}, S_{g}, R, C )$. Accordingly, parameter $g$ can be perceived as a control strategy $C$ responsible for mapping a set of symbols $\Gamma$ in accordance with a set of rules $R$. Variable $\bar{n}$ can be interpreted as an element of the set of initial states, \textit{i.e.} $i \in S_{i}$. The target condition $c$ can be understood as the set of goal states $S_{g}$. In addition, the unbounded minimization operator employs a parameter $m$ that represents the first argument where the target condition is met. Analogously, from a production system perspective, variable $m$ can be viewed as the first depth $d$ where a solution to the problem can be found. Finally, the condition $g(\bar{n},m) \neq c$ of the while loop is equivalent to applying the control strategy $C$ at total of $d$ times, \textit{i.e.} $C^{d}$, and evaluating if a goal state was reached.

\subsubsection{Iterative Search \label{sec:iterativeSearch}}

The fact that we are able to perform such mappings hints at the possibility of being able to develop our own quantum equivalent of the $\mu$-operator based on production system fundamentals. All that is required is a while loop structure, mimicking the iterative behaviour of the $\mu$-operator, that exhaustively examines every possibility for $d$ alongside $C$, until a goal state is found. Since we need to evaluate if applying $C^{d}$ leads to a solution we can combine the quantum production system oracle presented in Expression \ref{eq:complexValuedControlSystem2} alongside Grover's iterate for a total of $\sqrt{b^{d}}$ times in order to evaluate a superposition of all the available sequences of productions up to depth-level $d$, \textit{i.e.} $P_{R,d}$.  After applying Grover's algorithm, we can perform a measurement $M$ on the superposition, if the state $\xi$ obtained is a goal state, then the computation can terminate since a solution was found at depth $d$.

This process is illustrated in Algorithm \ref{algo:quantumIterativeDeepening} which receives as an argument a tuple $(\Gamma, i, S_{g}, R, C )$, where $i$ is an initial state, \textit{i.e.} $i \in S_{i}$. We choose to represent our procedure as a form of pseudocode that is in accordance with the conventions utilized in \cite{cormen2001}, namely: (1) indentation indicates block structure, \textit{e.g.} the set of instructions of the while loop that begins on line \ref{algo:quantumIterativeDeepening:whileLoopLine} consists of lines \ref{algo:quantumIterativeDeepening:whileLoopFirstLine}-\ref{algo:quantumIterativeDeepening:whileLoopLastLine}; (2) we use the symbol $\leftarrow$ to represent an assignment of a variable; and (3) the symbol $\triangleright$ indicates that the remainder of the line is a comment. 

\begin{algorithm}
\begin{algorithmic}[1]

\STATE \textbf{function }$f(\Gamma, i, S_{g}, R, C )$

	\STATE $ \qquad d \leftarrow 0$
	
	\STATE $ \qquad \xi \leftarrow \emptyset$
	
	\STATE $ \qquad \ket{s} \leftarrow i$
	
	\STATE $ \qquad \text{\textbf{while }} true$ \textbf{ do} \label{algo:quantumIterativeDeepening:whileLoopLine}
		
		\STATE $ \qquad \qquad \ket{p} \leftarrow \frac{1}{\sqrt{b^{d}}} \sum_{\forall x \in P_{R,d}} \ket{x}$ \label{algo:quantumIterativeDeepening:whileLoopFirstLine} \hspace{0.23cm}\COMMENT{Build superposition of productions} 
		
		\STATE $ \qquad \qquad \ket{h} \leftarrow \frac{1}{\sqrt{2}}\left(\ket{0} - \ket{1} \right)$ \label{algo:oracleApplication}
	
		\STATE $ \qquad \qquad \ket{\psi_{1}} \leftarrow C^{d} \ket{s} \ket{p} \ket{h}$  \label{algo:whileLoopLastLine:quantumRegister} \hspace{1.02cm}\COMMENT{Mark if goal states exist at depth $d$}
		
		\STATE $ \qquad \qquad \ket{\psi_{2}} \leftarrow G^{\sqrt{b^{d}}} \ket{\psi_{1}}$ \label{algo:applyGroversAlgorithm} \hspace{1.22cm}\COMMENT{Apply Grover's iterate}
		
		\STATE $ \qquad \qquad \xi \leftarrow M \ket{\psi_{2}}$ \hspace{2.04cm} \label{algo:performMeasurement} \COMMENT{Measure the superposition}
		
		\STATE $\qquad \qquad \text{\textbf{if }} \xi \in S_{g}$ \label{algo:quantumIterativeDeepening:stateMeasured}
		
		\STATE $\qquad \qquad \qquad \text{\textbf{return }} \xi$  \hspace{1.73cm}\COMMENT{If a goal state was found terminate}
		
		\STATE $\qquad \qquad \text{\textbf{else}}$
		
		\STATE $ \qquad \qquad \qquad d \leftarrow  d+1$ \label{algo:quantumIterativeDeepening:whileLoopLastLine} \hspace{1.65cm}\COMMENT{Otherwise, continue searching}
			
\end{algorithmic}
\caption{Quantum Iterative Deepening}
\label{algo:quantumIterativeDeepening}
\end{algorithm}

Line \ref{algo:oracleApplication} is responsible for applying the oracle alongside an initial state and all possible sequences of productions. Recall that register $\ket{h}$ will be set if goal states can be reached. Line \ref{algo:applyGroversAlgorithm} is responsible for applying Grover's algorithm. If goal states are present in the superposition, then Grover's amplitude amplification scheme allows for one of them to be obtained with probability $| \sin{ [ \frac{ \theta }{ 2 } ( \frac{ \pi }{ 2 } \sqrt{ \frac{ b^{d} }{ k } } +  1 ) ] } |^{2}$ \cite{nielsen2000}, where $k$ represents the number of solutions and $\theta = 2 \arccos{( \sqrt{ \frac{ b^{d} - k }{ b^{d} } })}$. It is possible that state $\ket{ \psi_{2} }$ contains a superposition of solutions. Therefore, measuring the system in Line \ref{algo:performMeasurement} will result in a random collapse amongst these. If the measurement returns an halt state, then register $\ket{p}$ will contain a sequence of productions leading to a goal state.  Once the associated sequence has been obtained one has only to apply each production of the sequence in order to determine precisely what was the goal state obtained \cite{tarrataca2011c} (Line \ref{algo:quantumIterativeDeepening:stateMeasured}). Otherwise, the search needs to be expanded to depth level $d+1$ and the production evaluation process repeated from the start. As a result, this procedure requires building a new superposition of productions $P_{R,d+1}$ each time a solution was not found in $P_{R,d}$.

Due to the probabilistic nature of Grover's algorithm there is also the possibility that the measurement will return a non halting state, even though $\ket{\psi_{2}}$ might have contained sequences of productions that led to goal states.  This issue can be circumvented to a certain degree. Notice that the sequences expressed by $P_{R,d+1}$  also contain the paths $P_{R,d}$ as subsequences. This means that when $P_{R,d+1}$ is evaluated the iteration procedure has the opportunity to re-examine $P_{R,d}$. As a result, operator $C$ would have the chance to come across the exact subsequences that had previously led to goal states but that were not obtained after the measurement. Therefore, the control strategy would need to be modified in order to signal an halt state as soon as a solution is found, \textit{i.e.} the shallowest production, independently of the sequence length being analyzed. With such a strategy the probability of obtaining a non-halting state in each unsought iteration level $d$ would be $1 - | \sin{ [ \frac{ \theta }{ 2 } ( \frac{ \pi }{ 2 } \sqrt{ \frac{ b^{d} }{ k } } +  1 ) ] } |^{2}$.

Each iteration of Algorithm \ref{algo:quantumIterativeDeepening} starts by building a superposition $\ket{p}$ spanning the respective depth level. This means that the original interference pattern that was possibly lost upon measuring the system in the previous iteration is rebuilt and properly extended by the tensor product that is performed with the new productions. Because of this process the computation is able to proceed as if undisturbed by the measurement. Such a reexamination comes at a computational cost which will be shown to be neglectable in Section \ref{sec:complexityAnalysis}. This behaviour contrasts with the original approach discussed by Deutsch where: (1) a computation would be applied to a superposition $\ket{\psi}$; (2) a measurement would eventually be made on the halt qubit collapsing the system to $\ket{\psi}^{\prime}$ and (3) if a goal state had not been obtained the computation would proceed with $\ket{\psi}^{\prime}$. 

\subsubsection{Complexity Analysis \label{sec:complexityAnalysis}}

Algorithm \ref{algo:quantumIterativeDeepening} represents a form of iterative deepening search, a general strategy employed alongside tree search, that makes it possible to determine an appropriate depth limit $d$, if one exists \cite{russell2003}. The first documented use of iterative deepening in the literature is in Slate and Atkin's Chess 4.5 program \cite{slate1977}, a classic application of an artificial intelligence problem. Notice that up until this moment we had not specified how to obtain a value for depth $d$, this was done deliberately since the essence of $\mu$-recursive functions relies in the fact that such a value may not exist. In general, iterative deepening is the preferred strategy when the depth of the solution is not known \cite{russell2003}.  Accordingly, the while loop will execute forever unless the state $\xi$ in line \ref{algo:quantumIterativeDeepening:stateMeasured}, obtained after the measurement, is a goal state. 

Since we employ Grover's algorithm we do not need to measure specifically the halting register. Instead it is possible to perform a measurement on the entire Hilbert space of the system in order to verify if a final state is obtained. This type of a control structure is responsible for guaranteeing the same type of partial behaviour that can be found on the classical $\mu$-operator. Consequently, Algorithm \ref{algo:quantumIterativeDeepening} also does not guarantee that variable $d$ will ever be found, \textit{i.e.} the search may not terminate. Line \ref{algo:whileLoopLastLine:quantumRegister} of our algorithm uses the register $\ket{r} = \ket{w} \ket{h} = \ket{s} \ket{p} \ket{h}$ described in Section \ref{sec:quantumProductionSystemOracle}.

Quantum iterative deepening search may seem inefficient, because each time we apply $C^{d}$ to a superposition spanning $P_{R,d}$ we are necessarily evaluating the states belonging to previous depth levels multiple times, $\forall d > 0$. However, the bulk of the computational effort comes from the dimension of the search space to consider, respectively $b^{d}$, which grows exponentially fast. As pointed out in \cite{korf1985} if the branching factor of a search tree remains relatively constant then the majority of  the nodes will be in the bottom level. This is a consequence of each additional level of depth adding an exponentially greater number of nodes. As a result, the impact on performance of having to search multiple times the upper levels is minimal. This argument can be stated algebraically by analysing the individual time complexities associated with each application of Grover's algorithm for the various depth levels. Such a procedure is illustrated in Expression \ref{eq:quantumIterativeSearchComplexity} which gives an overall time complexity of $O(\sqrt{b^{d}})$ remaining essentially unchanged from that of the original quantum search algorithm.

\begin{equation}
\sqrt{b^{0}} + \sqrt{b^{1}} + \sqrt{b^{2}} + \cdots + \sqrt{b^{d}} = O(\sqrt{b^{d}})
\label{eq:quantumIterativeSearchComplexity}
\end{equation}

By employing our proposal we are able to develop a quantum computational model with an inherent speedup relatively to its classical counterparts. Notice that this speedup is only obtained when searching through a search space with a branching factor of at least $2$ (please refer to \cite{tarrataca2010} \cite{tarrataca2011c}). In addition, if the set of goal states is defined to be the set of halt states, then we are able to use our algorithm to circumvent the halting problem. Our method is able to do so since it can compute a result without the associated disruptions of Deutsch's model. We employ such a term carefully, since it may be argued that the measurements performed during computation will inherently disturb the superposition. This is not a problem if a halt state is found. However, if such a goal state is not discovered, we move on to an extended superposition through $P_{R,d}$, representing an exponentially greater search space, where the states from the previous tree levels are included. Consequently, it becomes possible to recalculate the computation as if it had not been disturbed and without changing the overall complexity of the procedure. 

\section{Turing machine simulation \label{sec:turingMachineSimulation}}

The approach proposed in this work allows for the possibility of non-termination, without inherently interfering with the results of the quantum computation. This hints at the possibility that our approach can be applied to coherently simulate classical universal models of computation such as the Turing machine. Specifically, we are interested in determining what would be needed for our model of an iterative quantum production system to simulate any classical Turing machine?

We will begin by presenting a set of mappings between Turing machine concepts and production system concepts in a manner analogous to the trivial mapping described in \cite{franklin1997}. Both models employ some form of memory where the current status of the computation is stored. The Turing machine model utilises a tape capable of holding symbols. Each element of the tape can be referred to through a location. Tape elements are initially configured in a blank status, but their contents can be accessed and modified through primitive read and write operations. These operations are performed by a head that is able to address each element of the tape. As a result, the memory equivalent of the production system, respectively, the working memory should convey information regarding the current head position and the symbols, alongside the respective locations, on the tape. In addition, the tape employed in Turing's model has an infinite dimension. Consequently, the working memory must also possess an infinite character. 

The Turing machine model utilises a $\delta$ function to represent finite-state transitions. The $\delta$ functions maps an argument tuple containing the current state and the input being read to tuples representing a state transition, an associated output and some type of head movement. This set of transitions can be represented as a table whose rows correspond to some state and where each column represents some input symbol. Each table entry contains the associated transition tuple representing the next internal state, a symbol to be written, and a head movement. Notice, that this behaviour fits nicely into the fixed set of rules $R$ employed by production systems. Namely, $\delta$'s argument and transition tuples can be seen, respectively, as a precondition and associated action of a certain rule. Accordingly, for each table entry of the original Turing transition function it is possible to derive an adequate production rule, thus enabling the obtention of $R$. 

The only remaining issue resides in defining a control strategy $C$ that mimics the behaviour presented in Expression \ref{eq:complexValuedControlSystem2}. Consequently $C$ needs to choose which of the rules to apply by accessing the working memory, determining the element that is currently being scanned by the head, and establishing if a goal state is reached after applying some specific sequence of $R^{d}$ rules. Once this is done, we are able to apply our iterative quantum production system to simulate the behaviour of a classical Turing machine. The $\delta$-function conversion to an adequate database of productions is a simple polynomial-time procedure (please refer to \cite{abramsky1999} and \cite{sharma2006} for additional details). In addition, it is important to mention that this approach will only provide a speedup if the Turing machine simulated allows for multiple computational branches. Otherwise, if the computation is not capable of being parallelized then we gain nothing, performance-wise, from employing quantum computation.

\section{Conclusions \label{sec:conclusions}}

In this work we presented an approach for an iterative quantum production system with a built-in speedup mechanism and capable of the partial behaviour characteristic of $\mu$-recursive functions. Our proposal makes use of a unitary operator $C$ that can be perceived as mapping a total function since it maps for every possible input into a distinct output. However,  operator $C$ is employed in a quantum iterative deepening procedure that examines all path possibilities up to a depth level $d$ until a solution is found, if indeed there exists one. Due to the probabilistic nature of Grover's algorithm there is always the possibility that, upon measurement, a non-terminating state is obtained. As a consequence, the procedure would iterate to an additional level of productions and could therefore fail to recognize a halting state. This issue can be overcome through the development of specific control strategies capable of signaling that an halting state has been found at the shallowest production yielding such a conclusion, independently of the sequence length being analyzed.

Our model is able to operate independently of whether the computation terminates or not, a requirement associated with universal models of computation. As a result, it becomes possible for our model to exhibit partial behaviour that does not disturb the overall result of the underlying quantum computational process. This result is possible since: (1) Grover's algorithm effectively allows one to obtain halting states, if they exist, with high probability upon system observation; and (2) the overall complexity of this proposition remains the same of the quantum search algorithm. This procedure enables the development of verification-based universal quantum computational models, which are capable of coherently simulating classical models of universal computation such as the Turing machine.

\section*{Acknowledgements}

This work was supported by national funds through FCT – Fundação para a Ciência e a Tecnologia, under project PEst-OE/EEI/LA0021/2011 and FCT grant DFRH - SFRH/BD/61846/2009


\begin{thebibliography}{10}
\providecommand{\url}[1]{\texttt{#1}}
\providecommand{\urlprefix}{URL }
\expandafter\ifx\csname urlstyle\endcsname\relax
  \providecommand{\doi}[1]{doi:\discretionary{}{}{}#1}\else
  \providecommand{\doi}{doi:\discretionary{}{}{}\begingroup
  \urlstyle{rm}\Url}\fi
\providecommand{\bibAnnoteFile}[1]{%
  \IfFileExists{#1}{\begin{quotation}\noindent\textsc{Key:} #1\\
  \textsc{Annotation:}\ \input{#1}\end{quotation}}{}}
\providecommand{\bibAnnote}[2]{%
  \begin{quotation}\noindent\textsc{Key:} #1\\
  \textsc{Annotation:}\ #2\end{quotation}}
\providecommand{\eprint}[2][]{\url{#2}}

\bibitem{ekert1996}
Ekert A, Jozsa R (1996) Quantum computation and shor's factoring algorithm.
\newblock Rev Mod Phys 68: 733--753.

\bibitem{hilbert1900}
Hilbert D (1900) Mathematische probleme.
\newblock In: G{\"o}ttingen, editor, Proceedings of the International Congress
  of Mathematicians in Paris 1900. pp. 253-297.

\bibitem{lewis1981}
Lewis HR, Papadimitriou CH (1981) Elements of the Theory of Computation.
\newblock Upper Saddle River, NJ, USA: Prentice Hall PTR.

\bibitem{church1936a}
Church A (1936) A note on the entscheidungsproblem.
\newblock Journal of Symbolic Logic 1: 40--41.

\bibitem{turing1936}
Turing A (1936) On computable numbers, with an application to the
  entscheidungsproblem.
\newblock In: Proceedings of the London Mathematical Society. volume~2, pp.
  260-265.

\bibitem{deutsch1985}
Deutsch D (1985) Quantum theory, the church-turing principle and the universal
  quantum computer.
\newblock In: Proceedings of the Royal Society of London- Series A,
  Mathematical and Physical Sciences. volume 400, pp. 97-117.

\bibitem{hirvensalo2004}
Hirvensalo M (2004) Quantum Computing.
\newblock Berlin Heidelberg: Springer-Verlag.

\bibitem{bernstein1993}
Bernstein E, Vazirani U (1993) Quantum complexity theory.
\newblock In: STOC '93: Proceedings of the twenty-fifth annual ACM symposium on
  Theory of computing. New York, NY, USA: ACM, pp. 11--20.
\newblock \doi{http://doi.acm.org/10.1145/167088.167097}.

\bibitem{myers1997}
Myers JM (1997) Can a universal quantum computer be fully quantum?
\newblock Phys Rev Lett 78: 1823--1824.

\bibitem{ozawa1998a}
Ozawa M (1998) Quantum nondemolition monitoring of universal quantum computers.
\newblock Phys Rev Lett 80: 631--634.

\bibitem{ozawa1998b}
Ozawa M (1998) On the halting problem for quantum turing machines.
\newblock Technical report, Kyoto University, Japan.

\bibitem{linden1998}
{Linden} N, {Popescu} S (1998) {The Halting Problem for Quantum Computers}.
\newblock eprint arXiv:quant-ph/9806054 .

\bibitem{ozawa2002}
Ozawa M (2002) Halting of quantum turing machines.
\newblock In: Unconventional Models of Computation, Springer Berlin Heidelberg,
  volume 2509 of \emph{Lecture Notes in Computer Science}. pp. 58-65.
\newblock \doi{10.1007/3-540-45833-6_6}.
\newblock \urlprefix\url{http://dx.doi.org/10.1007/3-540-45833-6_6}.

\bibitem{miyadera2003}
{Miyadera} T, {Ohya} M (2003) {On Halting Process of Quantum Turing Machine}.
\newblock eprint arXiv:quant-ph/0302051 .

\bibitem{iriyama2004}
{Iriyama} S, {Ohya} M, {Volovich} I (2004) {Generalized Quantum Turing Machine
  and its Application to the SAT Chaos Algorithm}.
\newblock eprint arXiv:quant-ph/0405191 .

\bibitem{perdrix2004a}
{Perdrix} S, {Jorrand} P (2004) {Measurement-Based Quantum Turing Machines and
  their Universality}.
\newblock eprint arXiv:quant-ph/0404146 .

\bibitem{perdrix2004b}
{Perdrix} S, {Jorrand} P (2004) {Measurement-Based Quantum Turing Machines and
  Questions of Universalities}.
\newblock eprint arXiv:quant-ph/0402156 .

\bibitem{perdrix2006}
Perdrix S, Jorrand P (2006) Classically-controlled quantum computation.
\newblock Electronic Notes in Theoretical Computer Science 135: 119 - 128.

\bibitem{muller2007}
Muller M (2007) Quantum Kolmogorov Complexity and the Quantum Turing Machine.
\newblock Ph.D. thesis, Technical University of Berlin.

\bibitem{muller2008}
Muller M (2008) Strongly universal quantum turing machines and invariance of
  kolmogorov complexity.
\newblock Information Theory, IEEE Transactions on 54: 763 -780.

\bibitem{ross2010}
Duncan R, Perdrix S (2010) Rewriting measurement-based quantum computations
  with generalised flow.
\newblock In: Abramsky S, Gavoille C, Kirchner C, Meyer auf~der Heide F,
  Spirakis P, editors, Automata, Languages and Programming, Springer Berlin
  Heidelberg, volume 6199 of \emph{Lecture Notes in Computer Science}. pp.
  285-296.
\newblock \doi{10.1007/978-3-642-14162-1_24}.
\newblock \urlprefix\url{http://dx.doi.org/10.1007/978-3-642-14162-1_24}.

\bibitem{raussendorf2001}
Raussendorf R, Briegel HJ (2001) A one-way quantum computer.
\newblock Phys Rev Lett 86: 5188--5191.

\bibitem{raussendorf2003}
Raussendorf R, Browne DE, Briegel HJ (2003) Measurement-based quantum
  computation on cluster states.
\newblock Phys Rev A 68: 022312.

\bibitem{browne2011}
Browne D, Kashefi E, Perdrix S (2011) Computational depth complexity of
  measurement-based quantum computation.
\newblock In: Dam W, Kendon V, Severini S, editors, Theory of Quantum
  Computation, Communication, and Cryptography, Springer Berlin Heidelberg,
  volume 6519 of \emph{Lecture Notes in Computer Science}. pp. 35-46.
\newblock \doi{10.1007/978-3-642-18073-6_4}.
\newblock \urlprefix\url{http://dx.doi.org/10.1007/978-3-642-18073-6_4}.

\bibitem{perdrix2011}
Perdrix S (2011) Partial observation of quantum turing machines and a weaker
  well-formedness condition.
\newblock Electronic Notes in Theoretical Computer Science 270: 99 - 111.

\bibitem{mhalla2012}
{Mhalla} M, {Perdrix} S (2012) {Graph States, Pivot Minor, and Universality of
  (X,Z)-measurements}.
\newblock ArXiv e-prints .

\bibitem{abramsky1999}
Abramsky S, S A, Shore R, Troelstra A (1999) Handbook of computability theory.
\newblock Amsterdam, Netherlands: Elsevier.

\bibitem{post1943}
Post E (1943) Formal reductions of the general combinatorial problem.
\newblock American Journal of Mathematics 65: 197--268.

\bibitem{tarrataca2011b}
Tarrataca L, Wichert A (2012) A quantum production model.
\newblock Quantum Information Processing 11: 189-209.

\bibitem{stuart2004b}
Stuart T (2004) Partial recursive functions.
\newblock Technical report, University of Cambridge: Computer Laboratory:
  Faculty of Computer Science and Technology.

\bibitem{grover1996}
Grover LK (1996) A fast quantum mechanical algorithm for database search.
\newblock In: STOC '96: Proceedings of the twenty-eighth annual ACM symposium
  on Theory of computing. New York, NY, USA: ACM, pp. 212--219.
\newblock \doi{http://doi.acm.org/10.1145/237814.237866}.

\bibitem{grover2004a}
Grover LK, Radhakrishnan J (2004).
\newblock Is partial quantum search of a database any easier?
\newblock
  \urlprefix\url{http://www.citebase.org/abstract?id=oai:arXiv.org:quant-ph/0407122}.

\bibitem{kaye2007a}
Kaye PR, Laflamme R, Mosca M (2007) An Introduction to Quantum Computing.
\newblock USA: Oxford University Press.

\bibitem{brassard2000}
Brassard G, Hoyer P, Mosca M, Tapp A (2000).
\newblock Quantum amplitude amplification and estimation.
\newblock
  \urlprefix\url{http://www.citebase.org/abstract?id=oai:arXiv.org:quant-ph/0005055}.

\bibitem{chuang1998}
Chuang IL, Gershenfeld N, Kubinec M (1998) Experimental implementation of fast
  quantum searching.
\newblock Phys Rev Lett 80: 3408--3411.

\bibitem{tarrataca2011c}
Tarrataca L, Wichert A (2011) Problem solving and quantum computation.
\newblock Cognitive Computation 3: 510-524.

\bibitem{tarrataca2010}
Tarrataca L, Wichert A (2011) Tree search and quantum computation.
\newblock Quantum Information Processing 10: 475-500.

\bibitem{cormen2001}
Cormen TH, Leiserson CE, Rivest RL, Stein C (2001) Introduction to Algorithms,
  2/e.
\newblock MIT Press.

\bibitem{nielsen2000}
Nielsen MA, Chuang IL (2000) Quantum Computation and Quantum Information.
\newblock Cambridge, MA, USA: Cambridge University Press.

\bibitem{russell2003}
Russell SJ, Norvig P, Canny JF, Edwards DD, Malik JM, et~al. (2003) Artificial
  Intelligence: A Modern Approach (Second Edition).
\newblock Prentice Hall.

\bibitem{slate1977}
Slate D, Atkin LR (1977) Chess 4.5 - northwestern university chess program.
\newblock In: Chess Skill in Man and Machine. Berlin: Springer-Verlag, pp.
  82-118.

\bibitem{korf1985}
Korf RE (1985) Depth-first iterative-deepening : An optimal admissible tree
  search.
\newblock Artificial Intelligence 27: 97 - 109.

\bibitem{franklin1997}
Franklin S (1997) Artificial Minds.
\newblock MIT Press.

\bibitem{sharma2006}
Sharma A (2006) Theory of Automata and Formal Languages.
\newblock Laxmi Publications (P) Limited.
\newblock \urlprefix\url{http://books.google.pt/books?id=MKeF4o5o5JAC}.

\end{thebibliography}

\end{document}